# Enhancing Magnetic Ordering in Cr-doped Bi$_2$Se$_3$ using High-$T_C$ Ferrimagnetic Insulator

Wenqing Liu,[1,2] Liang He,[1,3] Yongbing Xu,[1,2] Koichi Murata,[3] Mehmet C. Onbasli,[4] Murong Lang,[3] Nick J. Maltby,[2] Shunpu Li,[2] Xuefeng Wang,[1] Caroline A. Ross,[4] Peter Bencok,[5] Gerrit van der Laan,[5] Rong Zhang,[1] Kang. L. Wang[3]

[1] York-Nanjing Joint Center for Spintronics and Nano Engineering (YNJC), School of Electronics Science and Engineering, Nanjing University, Nanjing 210093, China

[2] Spintronics and Nanodevice Laboratory, Department of Electronics, University of York, York YO10 5DD, UK

[3] Department of Electrical Engineering, University of California, Los Angeles, California 90095, USA

[4] Department of Materials Science and Engineering, Massachusetts Institute of Technology, Cambridge, Massachusetts 02139, USA

[5] Diamond Light Source, Didcot OX11 0DE, UK

**ABSTRACT**: We report a study of enhancing the magnetic ordering in a model magnetically doped topological insulator (TI), Bi$_{2-x}$Cr$_x$Se$_3$, via the proximity effect using a high-$T_C$ ferrimagnetic insulator Y$_3$Fe$_5$O$_{12}$. The FMI provides the TI with a source of exchange interaction yet without removing the nontrivial surface state. By performing the elemental specific X-ray magnetic circular dichroism (XMCD) measurements, we have unequivocally observed an enhanced $T_C$ of 50 K in this magnetically doped TI/FMI heterostructure. We have also found a larger (6.6 nm at 30 K) but faster decreasing (by 80% from 30 K to 50 K) penetration depth compared to that of diluted ferromagnetic semiconductors (DMSs), which could indicate a novel mechanism for the interaction between FMIs and the nontrivial TIs surface.

Three-dimensional TIs are insulating bulk materials that carry a conducting surface state, arising from the intrinsically strong spin-orbit interaction in the bulk band structure protected by time-reversal symmetry (TRS). While such unique systems offer nontrival surface states that can be utilized to perform dissipationless spin transport, it is equally important to break the TRS of TIs to realize novel physical phenomena. The newly discovered quantum anomalous Hall (QAH) effect,[1, 2] the hedgehog-like spin textures,[3] magnetoresistance switch effect,[4] carrier-independent long-range ferromagnetic order,[5] the predicted giant magneto-optical Kerr effect,[6] and magnetic monopole effect[7] are some of the fascinating examples.

Two categories of route for breaking TRS or introducing ferromagnetic order in TIs have been developed. One route is to dope the TI host with specific elements, by which ferromagnetism has been observed in Cr- and Mn-doped single crystals of $Sb_2Te_3$,[8, 9, 10] Fe-, and Mn-doped single crystals of $Bi_2Te_3$,[11, 12] and Mn- and Cr-doped thin films of $Bi_2Se_3$.[13, 14] The other routine is to engineer layered heterostructures, where the surface states of TIs experience the exchange interaction from an adjacent ferro- or ferri- magnetic material. This route subsequently can be divided into two ways in terms of ferro- or ferri- magnetic metal (FM) and ferro- or ferri- magnetic insulator (FMI) induction. Pioneering theoretical work[15, 16, 17] suggests that suitable FMIs have the potential to achieve a strong and uniform exchange coupling in contact with TIs without significant spin-dependent random scattering of helical carriers on magnetic atoms. Progresses are made experimentally in FMI/TI heterostructures including $GdN/Bi_2Se_3$ by Kandala *et al.*,[18] $EuS/Bi_2Se_3$ by Yang *et al.*,[19] and Wei *et al.*,[20] respectively, although the effect observed is limited to low temperature (< 22 K) due to the low $T_C$ of EuS. The interface magnetism of (anti-) FM/TI heterostructures, such as $Fe/Bi_2Se_3$,[21, 22, 23] $Co/Bi_2Se_3$,[22] and $Cr/Bi_2Se_3$[24] has also been investigated. Remarkably, Vobornik *et al.*[25] demonstrated that long-range ferromagnetism at ambient temperature can be induced in $Bi_{2-x}Mn_xTe_3$ by a deposited Fe overlayer. However, in the presence of a metallic layer, the nontrivial surface states of the TI can be significantly altered due to their hybridization with the bulk states of the (anti-) FM in contact. Besides, the metallic layer naturally short circuits the TI layer and therefore fundamentally restrict the device design.

Although magnetically doped TIs have demonstrated a pronounced capability with the magnetic proximity effect, very limited successful experimental demonstrations, especially by means of direct measurements like XMCD[19, 22, 25] have been reported. As shown in Figure 1 among all the building blocks within the research framework of FM or FMI/(magnetically doped) TI heterostructures, investigations of FMI/magnetically doped TI remains absent. Here, we present a work in realizing the proximity effect in an epitaxial $Bi_{2-}$ $_x$Cr$_x$Se$_3$/Y$_3$Fe$_5$O$_{12}$ heterostructure to fill up the void. Garnet-type Y$_3$Fe$_5$O$_{12}$ (YIG) is a well-known FMI with $T_C$ (~550 K) well above RT and a desirable large spin diffusion length. It contains two Fe ions occupying octahedral sites and three Fe ions occupying tetrahedral sites with opposite spin, resulting in ferrimagnetic ordering. The proximity effect has been demonstrated in PdPt/YIG,[26] Pt/YIG,[27] and Nb/YIG,[28] where interesting spin-transport properties were observed. While YIG-based heterostructures can consist of various materials, the best chance to realize strong exchange coupling may exist in the candidates with two-dimensional quantum surface states, such as TIs, as we have demonstrated in this report with a model TI/FMI heterostcrture Bi$_{2-x}$Cr$_x$Se$_3$/Y$_3$Fe$_5$O$_{12}$.

The 10 nm Bi$_{1.89}$Cr$_{0.11}$Se$_3$ thin films used in this study were grown in ultra-high vacuum using a Perkin-Elmer molecular-beam epitaxy (MBE) system on 50 nm YIG (111) film, which was pre-deposited on gallium gadolinium garnet (GGG) (111) substrate using pulsed-laser deposition (PLD).[29, 30] The X-ray diffraction (XRD) and magneto-optical Kerr effect (MOKE) magnetometry characterization of the YIG/GGG substrate have been published elsewhere.[31] High-purity Bi (99.9999%) and Cr (99.99%) were evaporated from conventional effusion cells at 470°C, while Se (99.99%) was formed from a cracker cell from SVTA at 240 °C, and the YIG/GGG (111) substrate was kept at 200 °C during growth. Interdiffusion of materials at the interface is not expected due to the high stability of YIG and the relatively low growth temperature of Bi$_{1.89}$Cr$_{0.11}$Se$_3$. 2 nm Al was then *in-situ* evaporated immediately after the growth of Bi$_{1.89}$Cr$_{0.11}$Se$_3$ to protect it from oxidation and environmental doping during transport to the synchrotron facility. Further details of the real-time reflection high-energy electron diffraction (RHEED) and the scanning transmission electron microscopy (STEM) characterization of the sample can be found in the supplementary materials.

The magnetic response of the epitaxial Bi$_{1.89}$Cr$_{0.11}$Se$_3$/YIG (111) thin film samples was first examined by the magneto-transport measurements by patterning into standard Hall bar devices, using conventional optical photolithography and a subsequent CHF$_3$ dry etching for 20 s. As shown in Figure 2A, six Hall channel contacts (10 nm Ti and 100 nm Au) were defined by e-beam evaporation. Standard four-terminal electrodes were fabricated to eliminate the contact resistance. A constant AC current of 0.05 ~ 0.1 µA with a frequency of 1300 Hz is fed through two outer contacts, and the voltage drop across the inner pads is measured to determine the resistance. By subtracting the ordinary Hall component, we plotted the anomalous Hall resistance ($R_{AHE} = R_{xy} - R_0 \cdot H$)[32] as a function of field applied perpendicularly to the film in Figure 2B. Non-zero $R_{AHE}$ was observable up to 50 K and vanished above 90 K. Figure 2C presents the temperature dependent $R_{AHE}$ of the Bi$_{1.89}$Cr$_{0.11}$Se$_3$ thin films on YIG, to which a Bi$_{1.89}$Cr$_{0.11}$Se$_3$ epitaxial thin film of the same thickness grown on highly resistive Si

(111) substrate was attached for comparison purpose. It can be seen that both these Cr-doped $Bi_2Se_3$ thin films exhibit Curie-like behavior, however, their magnetic ordering disappears at different temperatures, namely 30 K for $Bi_{1.89}Cr_{0.11}Se_3$/Si and up to 50 K for $Bi_{1.89}Cr_{0.11}Se_3$/YIG. The ferromagnetic ordering of $Bi_{1.89}Cr_{0.11}Se_3$/YIG was also observed from the field dependent longitudinal resistance ($R_{xx}$). In the low field region, weak anti-localization (WAL) with a clear cusp was observed at low temperatures, which is a characteristic feature associated with the gapped topological surface states below the critical temperature.[9,14,33] A typical $R_{xx}$ obtained at 3 K is presented in the inset of Figure 2D. The valleys of the WAL cusp exhibit a shift under the opposite field scanning directions, with the $R_{min}$ occurring approximately at the coercive field ($H_c$). We repeated the hysteretic longitudinal magnetic resistance measurement at elevated temperatures up to 90 K and found that $H_c$ remains observable till beyond 50 K, as plotted in Figure 2D, which is consistent with the $T_C$ estimated from $R_{AHE}$.

The element-specific technique of X-ray absorption spectroscopy (XAS) and XMCD at the Cr $L_{2,3}$ absorption edges were performed on beamline I10 at Diamond Light Source, UK, to probe the local electronic character of the magnetic ground state of the $Bi_{1.89}Cr_{0.11}Se_3$/YIG. Circularly polarized X-rays with ~100% degree of polarization were used in normal incidence with respect to the sample plane and parallel to the applied magnetic field, i.e., in Faraday geometry, as schematically shown in Figure 3A. XAS measurements were carried out at 6 - 300 K using total-electron yield (TEY) detection. XMCD was obtained by taking the difference of the XAS spectra, $\sigma^- - \sigma^+$, obtained by flipping the X-ray helicity at fixed magnetic field of 10 kOe, under which the sample is fully magnetized with little paramagnetic contribution.

A typical XAS and XMCD of the bilayer sample obtained by total-electron yield (TEY) at 6 K, normalized to the incident beam intensity, is presented in Figure 3C. The XAS spectra of Cr for both left- and right- circularly polarized X-rays show a white line at each spin-orbit split core level without prominent multiplet structure, except for a shoulder structure for the $L_2$ peak. The XAS spectral line shape resembles that of the ferromagnetic spinel-type Cr chalcogenides, i.e., $CdCr_2Se_4$, reported by Kimura *et al.*,[34] suggesting that the sample contains predominately $Cr^{3+}$ cations. Features of the obtained XMCD spectra are also in good agreement with those obtained for $CrFe_2O_4$ spinel ferrite powders, which can be well reproduced by multiplet calculations using a charge-transfer model with trivalent Cr cations on octahedral sites.[35] Consistent with the reported transport measurements,[14] the observed XAS and XMCD line shape also suggests that a majority of the Cr ions is incorporated within the crystal lattice by substituting onto the Bi sites (with a formal valance of $3^+$) in the TI matrix. The XAS and XMCD measurements were repeated at elevated temperatures and the dichroism at the Cr $L_3$ edge (575.3eV) was observable up to 50 K, as shown in Figure 3C, despite the decreasing intensity with increasing temperature. For clarity, the partial enlarged XAS of Cr $L_3$ edge at 30, 50, and 100 K, respectively, are presented in Figure 3B.

One of the most powerful aspects of the XMCD technique is that the average magnetic moment of the each element under interrogation can be quantitatively related to the integrated intensity of the XAS and XMCD spectra by applying the sum rules.[36, 37] Here, the orbital ($m_l$) and spin ($m_s$) moments of Cr were calculated according to equations

$$m_l = -\frac{4}{3} n_h \frac{\int_{L_{2,3}} (\sigma^- - \sigma^+) \, dE}{\int_{L_{2,3}} (\sigma^- + \sigma^+) \, dE} \quad (1)$$

$$m_s = -n_h \frac{6\int_{L_3} (\sigma^- - \sigma^+) \, dE - 4\int_{L_{2,3}} (\sigma^- - \sigma^+) \, dE}{\int_{L_{2,3}} (\sigma^- + \sigma^+) \, dE} \times SC - \langle T_z \rangle$$

where $E$, $n_h$, $SC$, and $\langle T_z \rangle$, respectively, represents the photon energy, the number of $d$ holes, the spin correction (SC) factor and the magnetic dipole term. In order to exclude the non-magnetic contribution of the XAS spectra an arctangent-based step function is used to fit the threshold.[38] The spectral overlap or $j$-$j$ mixing[36] was taken into account because of the relatively small spin-orbit coupling in the Cr $2p$ level. The value of SC, i.e. 2.0±0.2 for Cr, was estimated by calculating the $L_{2,3}$ multiplet structure for a given ground state, applying the sum rule on the calculated XMCD spectrum, and comparing the result with the spin moment calculated directly for this ground state.[39] Furthermore, $m_s$ needs to be corrected for the magnetic dipole term $\langle T_z \rangle$, however, its contribution is small for a Cr $t_{2g}^3$ configuration, giving an error < 5%.

Figure 4A-B presents the calculated $m_s$, $m_l$, and total magnetic moment ($m_{s+l}$) of Cr in $Bi_{1.89}Cr_{0.11}Se_3$/YIG bilayer at 6-300 K. Consistent with the magneto-transport results, the derived $m_{s+l}$ also exhibits a Curie-like behavior, pointing to a ferromagnetic phase of $Bi_{1.89}Cr_{0.11}Se_3$ at low temperatures. We obtained a remarkable $m_s$ = 1.38 ± 0.10 $\mu_B$/Cr and a small negative $m_l$ = -0.03 ± 0.02 $\mu_B$/Cr at 6 K. Noting that $m_s$ retains a sizable value of 0.58 ± 0.10 $\mu_B$/Cr at 30 K, we claim a pronounced increase of the $T_C$ in the $Bi_{1.89}Cr_{0.11}Se_3$ from 30 K, since otherwise $m_s$ should have nearly vanished at, or below, this point. With increasing temperature, $m_s$ reduces to 0.10 ± 0.10 $\mu_B$/Cr at 50 K, suggesting that the $Bi_{1.89}Cr_{0.11}Se_3$ is close to its $T_C$ here. Note that although the Fe dichroism in YIG remains sizably large up to RT (see supplementary materials), the Cr dichroism is no longer distinguishable from the noise at and above 100 K. As listed in Table 1, the reported $T_C$ of various kinds of magnetic TIs remained so far below ~30 K. Our demonstration of the ferromagnetic phase up to 50 K is significant in enhancing magnetic ordering of the magnetically doped TIs by the proximity effect with high-$T_C$ FMI, where the surface states of the TIs can be preserved with the insulating YIG.

The derived $m_l$ and $m_s$ have opposite signs, corresponding to antiparallel alignment of the spin and orbital moment in Cr. This agrees with the Hund's rule for trivalent Cr, whose 3$d$ shell is less-than-half full.[34] The octahedral crystal-field interaction quenches $m_l$, since the three $d$ electrons occupy the threefold degenerate majority-spin $t_{2g}$ orbitals, leading to a nearly vanishing $m_l$ as observed here. For similar reasons as for $m_l$, the magnetic-dipole term is small. Our observation of the total Cr magnetic moment is close to the value reported by Haazen *et al.*,[13] who performed superconducting quantum interference device (SQUID) measurements on a series of epitaxial $Bi_2Se_3$ with different Cr doping concentration (maximum $T_C$ = 20 K). In their work, the magnetic moment per Cr decreases significantly for $x > 5.2\%$, which coincides with a loss of $Bi_{2-x}Cr_xSe_3$ crystallinity. Such dependence is further evidence that the magnetization originates from the crystalline $Bi_{2-x}Cr_xSe_3$ phase. Cr clustering would not give rise to non-zero XMCD, since Cr is antiferromagnetic, as are the $Cr_xSe_y$ compounds, but therefore could have led to a reduced average Cr magnetic moment in the $Bi_{2-x}Cr_xSe_3$. However, since transport measurements are less sensitive to isolated ferromagnetic particles, our observed ferromagnetic behavior is still ascribed to be from the entire $Bi_{1.89}Cr_{0.11}Se_3$ thin film instead of magnetic clusters, if any.

Both the electrical transport and XMCD results point to the fact that between 30 and 50 K the magnetization of Cr can be attributed to the magnetic exchange coupling with YIG. We now address the ability of the YIG underlayer to induce magnetic ordering in $Bi_{1.89}Cr_{0.11}Se_3$ utilizing a model that was developed in the study of the proximity effect in dilute magnetic semiconductors (DMSs),[40, 41] as schematically sketched in Figure 4C. It is generally accepted that the XMCD intensity measured by TEY is attenuated by an exponentially decaying electron-escape probability, $\exp(-x/l_e)$,[42] we obtained

$$Cr\ \mathrm{XMCD} = \frac{\int_0^\infty \delta(x)\rho(x)e^{-x/\lambda_e}dx}{\int_0^\infty \rho(x)e^{-x/\lambda_e}dx} \qquad (2)$$

where $\lambda_e$ is the mean electron escape length. Provided (i) a sharp interface (see supplementary materials), (ii) a uniform distribution of Cr in $Bi_2Se_3$, and (iii) a steplike dichroism profile versus thickness $d$, we have $\delta(x) = \delta_{\exp}$ for $d_{YIG} < x < d_{\min}$ and $\delta(x) = 0$ elsewhere. Here $d_{\min}$ represents the lower limit, or the thickness of $Bi_{1.89}Cr_{0.11}Se_3$ contributing to the ferromagnetic signal at 30 - 50 K. Integration gives

$$d_{\min} = d_{Bi_{2-x}Cr_xSe_3} + \lambda_e \ln[\frac{\delta_{\exp}}{\delta_{sat}} + (1 - \frac{\delta_{\exp}}{\delta_{sat}})e^{-d_{Bi_{2-x}Cr_xSe_3}/\lambda_e}] \qquad (3)$$

Quoting the value $m_s$ = 1.38 $\mu_B$/Cr at 6 K, whose magnetic moment is considered to be intrinsic of the $Bi_{1.89}Cr_{0.11}Se_3$ without the effect of YIG, we obtain $\delta_{\exp}/\delta_{sat}$ = 43% and $d_{\min}$ = 6.6 nm at 30 K. This value quickly reduces to $d_{\min}$ = 1.8 nm at 50 K, where $\delta_{\exp}/\delta_{sat}$ = 7%. For the calculation of $d_{\min}$, we adopted $\lambda_e$ = 5 nm for the bulk mean electron escape length of $Bi_2Se_3$ and $d$ = 10 nm for the thin film thickness. Compared with the FM/DMS bilayer systems investigated by Maccherozzi *et al.*[40] using the same model, we have observed a larger penetration depth of the magnetic proximity effect at the interface of TI/FMI. The penetration depth decreases sharply with increasing temperature (i.e., > 80% from 30 K to 50 K). Typically, the proximity-induced magnetization in DMSs reduces only by ~10% within a comparable temperature range.[40] Such contrast may imply a unique type of interaction of the nontrivial surface state of TI with FMI. In other words, the penetration depth of the magnetic proximity effect in DMSs may have been limited by the contact barrier, while the conducting surface states of TIs may lift this limitation. It is generally believed that the origin of the proximity effect in a nonmagnetic/ferro- or ferri- magnetic (NM/FM) heterostructure arises from (spin-polarized) charge carriers propagating from the FM into the NM metal and vice versa, such that a finite spin polarization builds up close to the interface. A substantial reduction of such a spin polarization accumulation can be expected in a DMS, where charge carriers can hardly penetrate. In contrast, our results suggest that such charge carrier propagation can be less suppressed in the TI/FMI system due to the presence of the conducting surface states, though whose ability is more sensitive to the temperature variation.

To summarize, we have observed strongly dichroic XAS spectra of Cr in $Bi_{1.89}Cr_{0.11}Se_3$/YIG up to 50 K, corresponding to an enhanced $T_C$ in this magnetically doped TI/FMI exchange system. The unique elemental selectivity of the XMCD technique has enabled a direct determination of the proximity-induced magnetization of $Bi_{1.89}Cr_{0.11}Se_3$ at its interface with YIG. We have found a larger but faster dropping penetration depth in such a magnetic TI/FMI heterostructure compared to that in DMS/FM, which may be due to a novel mechanism of the interaction of FMIs and nontrivial TIs surface. The result is furthermore the first demonstration of XMCD in Cr-doped $Bi_2Se_3$ epitaxial thin films, presenting an unambiguous picture of the electronic and magnetic state of the magnetic dopants in the TIs. Our study takes an important step towards realizing TI-based spintronics. Future work to explore the coupling mechanism of the TI/FMI interface and its dependence on the band filling will help to find experimental approaches to further increase the $T_c$ in the magnetically doped TI/FMI hybrid material systems, which has strong implications for both fundamental physics and emerging spintronics technology.

**ASSOCIATED CONTENT**

Supplementary materials accompany this paper.

**AUTHOR INFORMATION**

**Corresponding Authors**

*E-mail: yongbing.xu@york.ac.uk.

*E-mail: rzhang@nju.edu.cn.

*E-mail: wang@seas.ucla.edu.

**Author Contributions**

Wenqing Liu and Liang He contributed equally to this work.

**Notes**

The authors declare no competing financial interest.

**ACKNOWLEDGEMENT**

This work is supported by the State Key Programme for Basic Research of China (Grants No. 2014CB921101), NSFC (Grants No. 61274102), UK STFC, DARPA Meso program under contract No.N66001-12-1-4034 and N66001-11-1-4105. We thank Y. Wang of Zhenjiang University for the help with the TEM characterization. C. A. Ross and M. C. Onbasli acknowledge support of FAME, a STARnet Center of DARPA and MARCO, and by the NSF. Diamond Light Source is acknowledged for beamtime on I10.

**TABLES**

| System | $T_C$ | Ref. |
|---|---|---|
| $Bi_{2-x}Cr_xSe_3$/YIG | 50 K | [*] |
| $Bi_{2-x}Cr_xSe_3$ | 20 K | [13] |
| $Sb_{2-x}Cr_xTe_3$ | 20 K | [8] |
| $Cr_x(Bi_ySb_{1-y})_2Te_3$ | 11 K | [9] |
| $Bi_{2-x}Mn_xTe_3$ | 12 K | [11] |
| $Bi_{2-x}Fe_xTe_3$ | 12 K | [12] |

Table 1. The $T_C$ of the magnetically doped TIs from the literatures. Typical values of $T_C$ from reports, which remains generally below ∼30 K till today. Our work [*] has demonstrated an enhanced $T_C$ up to ∼50 K via the proximity effect using a YIG under layer.

**FIGURES**

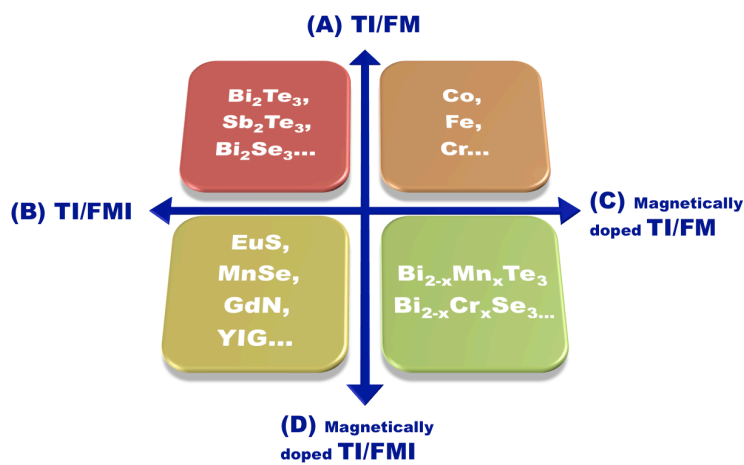

Figure 1. Schematic diagram of the research framework of magnetic TI hybrid systems. Squares representing FM (top right) and FMI (bottom left) integrate with those representing TI (top right) and magnetically doped TI (bottom right), encompassing four categories of subjects of magnetic TI by engineering layered heterostructures, namely, investigations of (A) TI/FM including Fe/$Bi_2Se_3$,[21] Co/$Bi_2Se_3$,[22] and Cr/$Bi_2Se_3$;[24] (B) TI/FMI including MnSe/$Bi_2Se_3$,[15, 16] GdN/$Bi_2Se_3$,[18] EuS/$Bi_2Se_3$;[19, 20] (C) doped TI/ FM Fe/$Bi_{2-x}Mn_xTe_3$;[25] and (D) doped TI/FMI, the remaining unexplored area, on which this letter reports on.

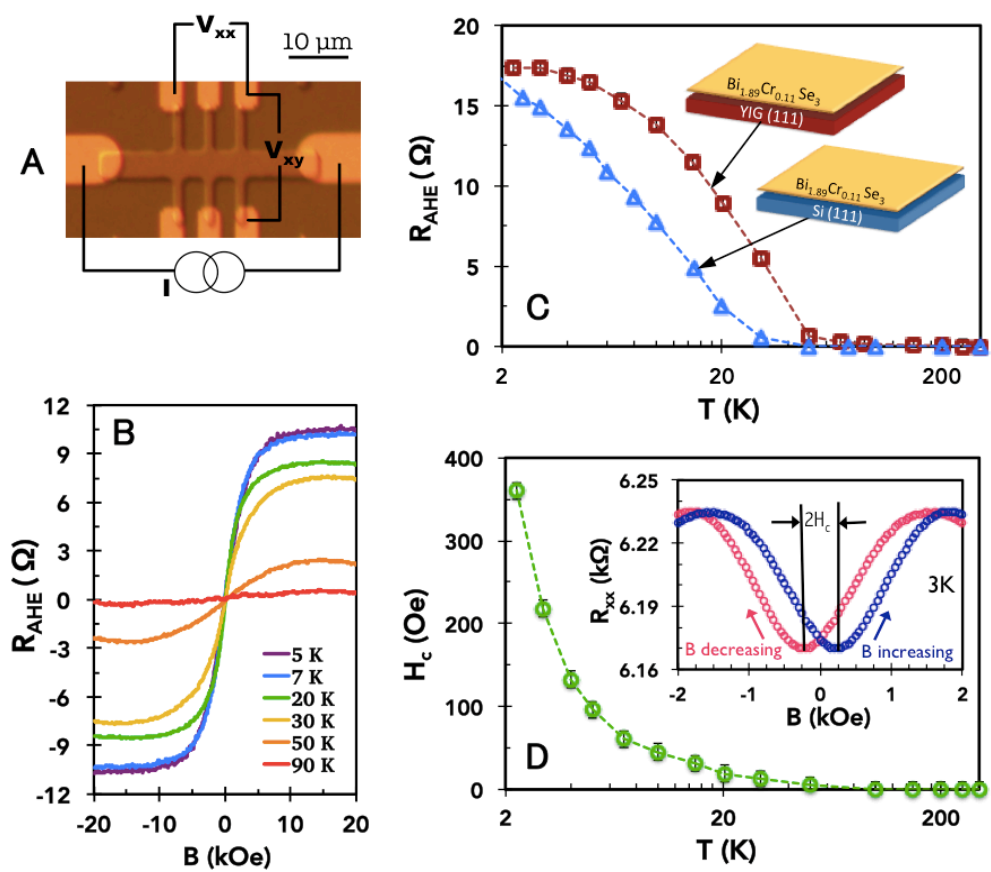

Figure 2. Magneto-transport measurement. (A) Schematic diagram of the experimental set up for the transport measurements. (C) AHE of the $Bi_{1.89}Cr_{0.11}Se_3$/YIG thin film versus magnetic field at 20-90 K. (B) Comparison of the AHE versus temperature of the $Bi_{1.89}Cr_{0.11}Se_3$ thin films grown on YIG (111) and Si (111), respectively. (D) The $H_c$ of $Bi_{1.89}Cr_{0.11}Se_3$/YIG versus temperature. Inset: the shift of the valleys of WAL cusp obtained at 3 K, associated with the $H_c$. The arrows represent the scanning direction of the magnetic field.

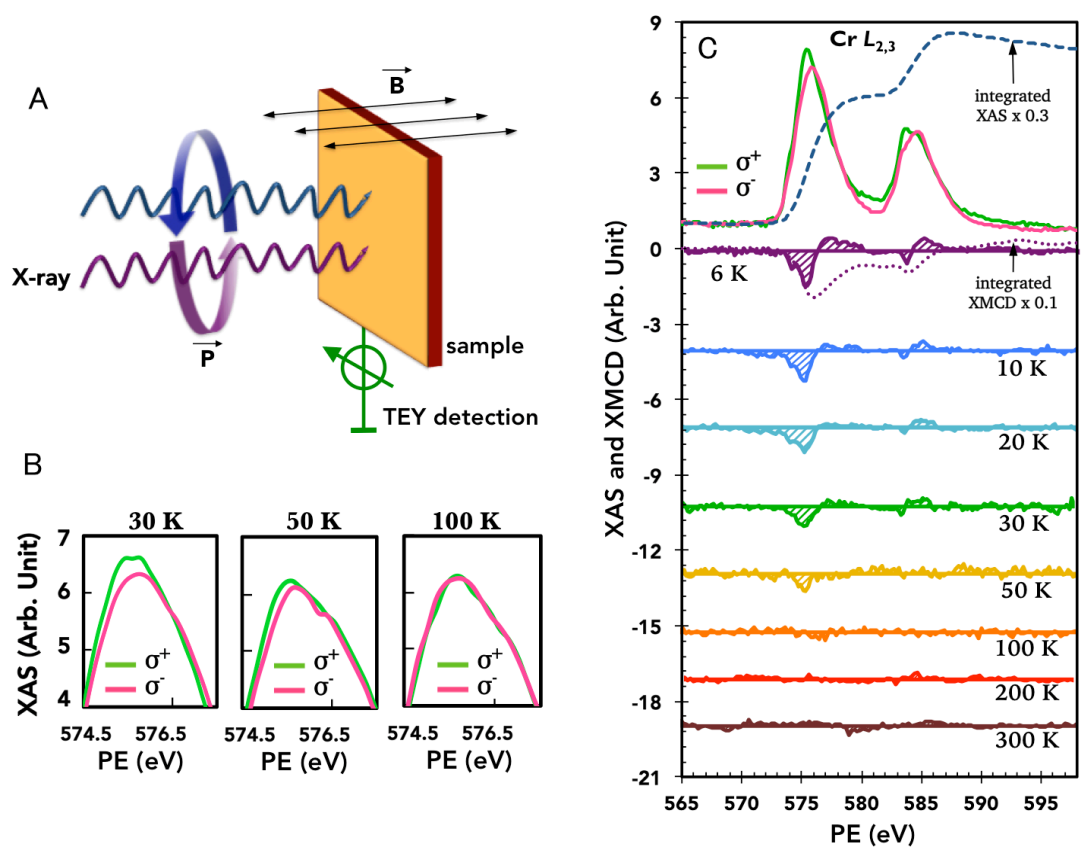

Figure 3. XAS and XMCD measurement. (A) Schematic diagram of the experimental set up of the XMCD experiment. (B) The partial enlarged XAS of Cr $L_3$ edge at 30, 50, and 100 K, respectively. (C) Typical pair of XAS and XMCD spectra of the $Bi_{1.89}Cr_{0.11}Se_3$/YIG bilayer sample obtained at 6 K and their integrals (the XAS and XMCD spectra are offset for clarity) and that at elevated temperatures, where dichroism at the Cr $L_3$ edge (spectra at different temperatures are offset vertically for clarity). The dash lines indicate the integration of the spectra.

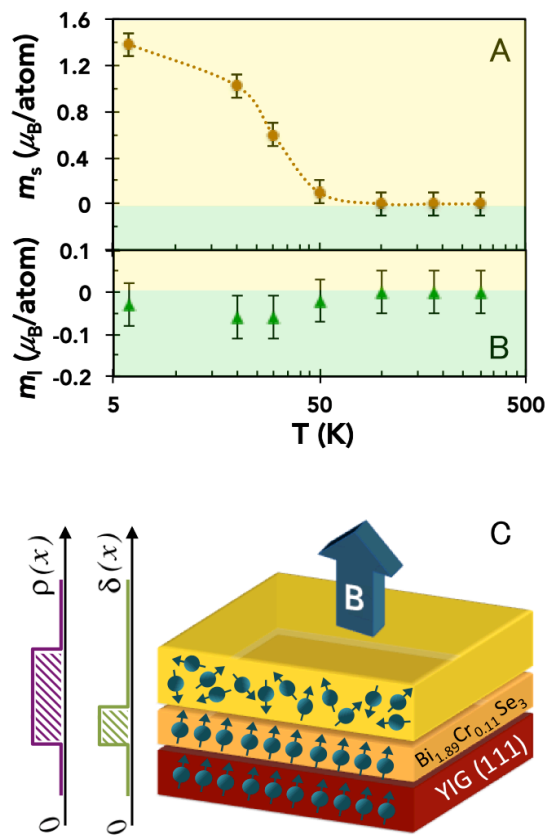

Figure 4. The enhanced magnetic ordering of $Bi_{1.89}Cr_{0.11}Se_3$ via the proximity effect. (A)-(B) The $m_s$ and $m_l$ of Cr at 6-300 K derived from the sum rules. The dashed line is a guide to the eye. (b) A schematic diagram of the model used to estimate the proximity length, showing the Cr distribution $\rho(x)$ and the ferromagnetically ordered Cr distribution $\delta(x)$ at given temperature, as described in the text.